\documentclass[aip,jcp,10pt,twocolumn,superscriptaddress,nofootinbib,reprint]{revtex4-1}

\usepackage{bm}
\usepackage[english]{babel}
\usepackage{graphicx}
\usepackage{hyperref}
\usepackage[utf8]{inputenc}
\usepackage{mathtools}
\usepackage[expansion=true,protrusion=true]{microtype}
\usepackage{siunitx}
\usepackage{textcomp}
\usepackage{textgreek}
\usepackage[usenames,dvipsnames]{xcolor}

\DeclareSIUnit\Molar{M}
\hypersetup{hidelinks}
\usepackage{cleveref}

\renewcommand{\eqref}[1]{(Eq.\,\ref{#1})}
\crefformat{equation}{(Eq.\,#2#1#3)}

\newcommand{\lD}{\lambda_{D}}
\newcommand{\kT}{k_{B}T}
\newcommand{\tens}[1]{\bar{\bm{#1}}}
\let\vec\bm

\usepackage{natbib}

\begin{document}

\title{Nanoparticle Translocation through Conical Nanopores:\protect\\ A Finite Element Study of Electrokinetic Transport}

\author{Georg Rempfer}
\email{georg.rempfer@icp.uni-stuttgart.de}
\author{Sascha Ehrhardt}
\author{Christian Holm}
\affiliation{Institute for Computational Physics (ICP), University of Stuttgart, Allmandring 3, 70569 Stuttgart, Germany}
\author{Joost de Graaf}
\altaffiliation{Current address: School of Physics and Astronomy, University of Edinburgh, Scotland, Edinburgh EH9 3JL, United Kingdom.}
\affiliation{Institute for Computational Physics (ICP), University of Stuttgart, Allmandring 3, 70569 Stuttgart, Germany}

\date{\today}

\begin{abstract}
Recent years have seen a surge of interest in nanopores because such structures show a strong potential for characterizing nanoparticles, proteins, DNA, and even single molecules. These systems have been extensively studied in experiment as well as by all-atom and coarse-grained simulations, with a strong focus on DNA translocation. However, the equally interesting problem of particle characterization using nanopores has received far less attention. Here, we theoretically investigate the translocation of a nanoparticle through a conical nanocapillary.  We use a model based on numerically solving the coupled system of electrokinetic continuum equations, which we introduce in detail. Based on our findings, we formulate basic guidelines for obtaining the maximum current signal during the translocation event, which should be transferable to other nanopore geometries. In addition, the dependence of the signal strength on particle properties, such as surface charge and size, is evaluated. Finally, we identify conditions under which the translocation is prevented by the formation of a strong electro-osmotic barrier and show that the particle may even become trapped at the pore orifice, without imposing an external hydrostatic pressure difference.
\end{abstract}

\maketitle

The use of microfluidic structures to characterize colloidal particles, by measuring the spike in current as the particle moves through a channel, has a long history, dating back to the simple Coulter counter for red blood cells.~\cite{h_means_1953, Fraikin11, Kozak11, Luo14, Weatherall15} This phenomenon is well-understood in the simple channel geometry and known to be related to the exclusion of ions from the conduit leading to a lower current,~\textit{i.e.}, a modification of the resistance. Breakthroughs in fabrication have pushed the boundaries of traditional microfluidic devices to the nanoscale. Nanopores, especially, have received considerable attention because they are often envisioned to enable low-cost DNA sequencing.~\cite{Zwolak08} Understanding the principles of DNA translocation through a nanopore using theoretical and numerical modeling is, sadly, exceptionally challenging.~\cite{heng_sizing_2004, aksimentiev_microscopic_2004, heng_beyond_2005, aksimentiev_imaging_2005, storm_fast_2005, smeets_salt_2006, stein_electrokinetic_2010, kesselheim_applying_2011, kesselheim_effects_2012, laohakunakorn_dna_2013, kesselheim_origin_2014, rempfer_selective_2016} In addition, fully accurate sequencing of large strands of DNA has not been achieved to date, which raises questions concerning the feasibility of pore-based characterization.~\cite{mikheyev_first_2014}

However, conduit current measurements of nanoparticles translocating through (conical) nanopores have recently been experimentally achieved and theoretically modeled.~\cite{lan_nanoparticle_2011, lan2011pressure, li_single_2013, Luo14, lan_effect_2014, tsutsui_particle_2016} This problem also appears simpler than molecular translocation, as it can be investigated by solving the coupled system of electrokinetic (Stokes, Poisson, and Nernst-Planck) equations with appropriate boundary conditions. These equations describe fluid flow and ionic current through the nanopore and around the nanoparticle. It was found that electrokinetic simulations capture experimental results on translocation of nanoparticles well. This has led to further investigation, in which the sensitivity of the modulations in the ionic current conducted by the pores during the translocation process was exposed.~\cite{lan_nanoparticle_2011, lan_effect_2014, tsutsui_particle_2016} Finally, means by which to control the translocation process have been investigated.~\cite{german_controlling_2013} Unfortunately, the complicated geometrical problem posed by a nanoparticle translocating through a nanopore makes it difficult to get a handle on the problem analytically and generally one must resort to numerical methods.

Even the simpler problem of pure electroosmotic flow (EOF;~\textit{i.e.}, without a DNA or nanoparticle) through a nanopore can only be analytically tackled in highly simplified geometries using rough approximations.~\cite{Laohakunakorn15a,Laohakunakorn15b} Recent experimental studies have shown that even these relatively simple systems comprised of only a conical nanopore display rich behavior when they are immersed in a saline solution and an electric field is applied. Namely, these asymmetric pores have the ability to rectify current and flow and are a fluid-dynamic analogue of a diode.~\cite{Woerman03,Siwy04,Siwy06,Laohakunakorn15a,Laohakunakorn15b} Numerical approaches and limited theoretical work have proven quite successful in capturing the experimental trends and furthering understanding of the underlying physical principles.~\cite{Ai10,Mao13,Mao14,Laohakunakorn15a,Laohakunakorn15b}

In this paper, we solve the electrokinetic equations numerically via the finite-element method, to investigate the problem of current and flow rectification in a conical nanopore, as well as the translocation of a nanoparticle though such a pore. We thus remain in the regime, for which previous studies have shown the electrokinetic equations to perform well in describing experiments, but we go beyond the established literature in several ways. We use improved numerical approaches and meshing to allow for accurate simulation of large nanopores, as \emph{e.g.}\ considered in the Keyser group.~\cite{thacker_studying_2012, li_single_2013} In addition, we identify environmental conditions favorable to nanoparticle translocation through such nanopores and parameters which lead to other behaviors, such as trapping and repulsion.

Our main results are the following. We find that the observed currents and flow have a significant finite-size effect, which asymptotically scales with the inverse length of the capillary. To faithfully represent the experimental systems with capillary lengths in the \si{\milli\m} to \si{\centi\m} range requires simulation of a \SI{20}{\micro\m} capillary and \SI{100}{\micro\m} of the surrounding reservoir. We also investigate the influence of advective transport of ions. We find that for low salt concentration only there is a noticeable influence on the current and flow through the capillary, but in that regime the modification of the physical observables is less than \SI{10}{\percent}. To achieve this result, we required our enhanced meshing, as well as a modified forcing term for the fluid.~\cite{rempfer_limiting_2016} Our result also justifies neglecting of advection in previous works, \emph{e.g.},~\cite{Laohakunakorn15b} where the lower-resolution meshing used would have made it difficult to accurately assess its influence.

We also recover the result that the local ionic excess near the nanopore tip is closely related to the emergence of rectification. That is, the polarization effect first observed by Laohakunakorn~\emph{et al.}~\cite{Laohakunakorn15b} We further study the effect of salt concentration (or equivalently the Debye length), on the rectification ratio that can be achieved by the conical nanopore. The optimum ionicity for achieving rectification is identified and the various limiting behaviors are discussed. This optimum could prove relevant to tuning the sensitivity of this nanopore system for particle characterization.

Our observations for the pure EOF system, as well as their correspondence to literature results, gave us confidence applying our numerical scheme to study the translocation of a nanoparticle through the nanopore. This part of our investigation is in a similar vein as the work by Refs.~\cite{lan_nanoparticle_2011, li_single_2013, lan_effect_2014}, where the translocation of a nanoparticle was considered along the symmetry axis of the pore. However, we do not consider an externally imposed pressure difference, as in Ref.~\cite{german_controlling_2013, Luo14, lan2011pressure} We find that the translocation event has a measurable impact on the current through the pore --- up to \SI{50}{\percent} compared to the base current. Similar nanoparticle translocation effects have been numerically and experimentally considered by the group of White~\cite{lan_nanoparticle_2011, lan_effect_2014} and we obtain comparable signal shapes. 

We show that to achieve the strongest translocation signal, a high salt concentration should be used in combination with an electric field that points into the bulk out of the pore orifice. High salt and an oppositely directed field leads to a much weaker signal. Despite not having an external pressure difference,~\cite{Luo14,german_controlling_2013,lan2011pressure} we find that translocation is not guaranteed. High salt concentrations allow for unimpeded translocation through the nanopore. For low salt concentrations, the sphere may become trapped close to the pore orifice, or even strongly repelled from the orifice, preventing translocation. For high salt and favorable field direction, the pore system shows remarkable sensitivity to the size and charge of the sphere, which we also characterize, and that are in line with the observation by Lan~\textit{et al.}~\cite{lan_nanoparticle_2011}

The results presented in this paper thus provide new insight into the simulation of conical nanopore geometries using the electrokinetic equations via the finite-element method. In particular, our method of meshing and forcing~\cite{rempfer_limiting_2016} should prove instrumental in limiting numerical artifacts. These two improvements allow for the precise study of nanoparticle translocation events through conical (and other) nanopores. Interestingly, translocation is not necessarily guaranteed and depends sensitively on the applied voltage and bulk ionicity of the system. It might be possible to recover these findings in an experimental system. Future studies will focus on extending our results to off-axis calculations for the nanoparticle translocation, as well as imposing external hydrodynamic pressure differences over the nanopore, as these have been shown to strongly impact the translocation event.

The remainder of this manuscript is structured as follows. We first introduce the mathematical model by which we describe nanoparticle transport through nanopores, namely the electrokinetic equations in~\Cref{sec:ekin_eq}. We follow up in~\Cref{sec:fem} with a thorough discussion of the finite-element method (FEM) employed to solve the electrokinetic equations in the complex geometry formed by a glass nanocapillary. In~\Cref{sec:rectification} we demonstrate that our FEM model correctly reproduces the current and flow rectification effects observed in these nanocapillary systems. Next, we apply the FEM model to investigate the translocation of a colloidal sphere through the nanocapillary in~\Cref{sec:translocation}. We give conclusions and present an outlook in~\Cref{sec:conclusion}.

\section{\label{sec:ekin_eq}The Electrokinetic Equations}

In this section we provide details of our physical modeling of fluid flow and current rectification in the conical nanopore geometry. We also describe the way in which we simulate the translocation of a nanoparticle through such a pore. The physics of all of these phenomena can be described by the electrokinetic equations, comprised of the coupled system of Stokes, Poisson, and Nernst-Planck equations, which we introduce first. The specific problems of rectification by nanopores and nanoparticle translocation through them amount to a set of boundary conditions on the electrokinetic equations, which are discussed next.

\subsection{Bulk Equations}

Let us now turn to the electrokinetic equations, without concerning ourselves with boundary conditions. We model the dynamics of the water separately from the dissolved ionic species and use the incompressible Stokes equations to model the flow of water:
\begin{equation}
\label{eq:stokes}
\begin{aligned}
\eta \bm{\nabla}^{2} \vec{u} &= \nabla p - \vec{f} ,\\
\nabla \cdot \vec{u} &= 0 .
\end{aligned}
\end{equation}
The incompressible Stokes equations are valid for flow of a Newtonian fluids of shear viscosity $\eta$ at low Reynold's numbers, for which viscous forces dominate over inertial forces. Here $p$ denotes the hydrostatic pressure, and $\vec f$ is a yet to be determined coupling force, coupling the fluid motion to that of the dissolved ions.

In the following, we closely follow our original derivation in Ref.~\cite{rempfer_limiting_2016}, which can be consulted for additional details by the interested reader. We model two oppositely charged ionic species, indicated in the following by the $\pm$ subscripts. The dynamics of these dissolved ionic species relative to the underlying fluid motion can be modeled as diffusive. This allows one to derive the ionic fluxes in a reference frame co-moving with the local fluid velocity $\vec u$ from a free energy under the local equilibrium approximation. Equivalently to Poisson-Boltzmann theory, the free energy density in the electrokinetic equations consists of an ideal-gas part and a contribution to the internal energy coming from electrostatic interactions:
\begin{equation}
\begin{multlined}
\label{eq:free_energy}
f \big( \{c_{\pm}(\vec{r})\} \big) = \underbrace{ \textstyle\sum_{\pm} \kT c_{\pm}(\vec{r}) \left[ \log \left\lbrace \Lambda_{\pm}^{3} c_{\pm}(\vec{r}) \right\rbrace - 1 \right] }_{\mathrm{ideal~gas~contribution}} \\
+ \underbrace{z_{\pm} e c_{\pm}(\vec{r}) \Phi(\vec{r})}_{\mathclap{\mathrm{electrostatic~contribution}}} .
\end{multlined}
\end{equation}
Here, $c_{\pm}$ denotes the ionic densities, $\Lambda_{\pm}$ their thermal de Broglie wavelengths, $z_{\pm}$ their valencies, and $\Phi(\vec{r})$ the electrostatic potential.

The chemical potential of the two ionic species is given by
\begin{equation}
\label{eq:chempot}
\mu_{k}(\vec{r}) = \frac{\delta f(c_{k})}{\delta c_{k}} = \kT \log(\Lambda_{k}^{3} c_{k}(\vec{r})) + z_{k} e \Phi(\vec{r}) .
\end{equation}
The above expression finally allows one to derive a first-order approximation to the thermodynamic driving force as the gradient of the chemical potential~\eqref{eq:chempot} and use this driving force to formulate an expression for the diffusive flux:
\begin{equation}
\begin{aligned}
\label{eq:j_diff}
\vec{j}_{\pm}^{\text{diff}} &= \mu_{\pm} c_{\pm} \left( -\nabla \mu_{\pm} \right) = -D_{\pm} \nabla c_{\pm} - \mu_{\pm} z_{\pm} e c_{\pm} \nabla \Phi . 
\end{aligned}
\end{equation}
Here,  $D_{\pm}$ and $\mu_{\pm}$ denote the diffusion coefficient and the mobility of the different ionic species, which are related by the Einstein-Smoluchowski relation $D_{\pm} / \mu_{\pm} = \kT$.~\cite{einstein1905a, smoluchowski1906a}

Expressing the ionic fluxes in a reference frame at rest results in an additional contribution due to the advection of ions with the fluid. The final expression for the total ionic fluxes assumes the shape of a diffusion-advection equation and reads:
\begin{equation}
\label{eq:flux}
\vec{j}_{\pm} = \underbrace{-D_{\pm} \nabla c_{\pm} - \mu_{\pm} z_{\pm} e c_{\pm} \nabla \Phi}_{\vec{j}_{\pm}^{\text{diff}}} + \underbrace{c_{\pm} \vec{u}}_{\vec{j}_{\pm}^{\text{adv}}} .
\end{equation}
For systems with characteristic length scales in the nanometer range, the flux is completely dominated by the diffusive contribution to $\vec{j}_{\pm}^{\text{diff}}$, which allows one to neglect the advective contribution to $\vec{j}_{\pm}^{\text{diff}}$. This approximation corresponds to the low-P{\'e}clet number limit. The P{\'e}clet number arises in the de-dimensionalized form of the diffusion-advection equation~\eqref{eq:flux} and quantifies the magnitude of the advective flux $\vec j_\pm^\text{adv}$ relative to the diffusive flux $\vec j_\pm^\text{diff}$. We investigate the quality of this approximation for our model system in~\Cref{sec:rectification}.

In the absence of sources or sinks for the ionic species, the ionic fluxes must follow the continuity equation:
\begin{equation}
\label{eq:continuity}
\partial_{t} c_{\pm} = -\nabla \cdot \vec{j}_{\pm} .
\end{equation}
In stationary situations, none of the fields vary over time ($\partial_{t} = 0$), and \cref{eq:flux} and \cref{eq:continuity} can be combined into
\begin{equation}
\label{eq:concentrations}
\nabla \cdot (D_{\pm} \nabla c_{\pm} + \mu_{\pm} z_{\pm} e c_{\pm} \nabla \Phi - c_{\pm} \vec{u}) = 0 .
\end{equation}

Neglecting magnetic effects, the electrostatic potential $\Phi$ fulfills the Poisson equation:
\begin{equation}
\label{eq:poisson}
\nabla \cdot (\varepsilon \nabla \Phi) = -\varrho = -\textstyle\sum_{\pm} z_{\pm} e c_{\pm} .
\end{equation}
Here the charge density $\varrho$ is given in terms of the ionic species concentrations $c_{\pm}$. The permittivity $\varepsilon = \varepsilon_{0} \varepsilon_r(\vec{r})$ is the product of the vacuum permittivity $\varepsilon_{0}$ and the local relative permittivity $\varepsilon_r(\vec{r})$ of the medium.\\

Having defined all other quantities, we can now come back to the force density $\vec f$, coupling the fluid motion to the motion of dissolved ions from \Cref{eq:stokes}. In a previous investigation,~\cite{rempfer_limiting_2016} we determined that so-called spurious fluxes and spurious flow from the dominating errors in numerical solutions of the electrokinetic equations. To reduce these artifacts, we developed a modified fluid coupling deviating from the approach typically taken in the literature,~\cite{hsu_diffusiophoresis_2010, rice_electrokinetic_1965, burgreen_electrokinetic_1964, laohakunakorn_electroosmotic_2015, laohakunakorn_electroosmotic_2015-1, obrien_electrophoretic_1978, berg_exact_2009, white_ion_2008, daiguji_nanofluidic_2005, hlushkou_numerical_2005, dhopeshwarkar_transient_2008} which consists of a formulation based only on electrostatic forces. We include the ideal gas pressure of the ionic species into the fluid coupling force:
\begin{equation}
\label{eq:coupling}
\vec{f} = -\textstyle\sum_{\pm} ( \kT \nabla c_{\pm} + z_{\pm} e c_{\pm} \nabla \Phi )
\end{equation}
We demonstrated analytically and with numerical simulations that this modified coupling results in the same solutions for the flow field, but reduces spurious flow artifacts by several orders of magnitude.~\cite{rempfer_limiting_2016}

Summarizing, the stationary electrokinetic equations are given by the following system of equations:
\begin{equation}
\begin{aligned}
\label{eq:summary}
0 &= \nabla \cdot (D_{k} \nabla c_{k} + \mu_{k} z_{k} e c_{k} \nabla \Phi - c_{k} \vec{u}) ,\\
\nabla \cdot (\varepsilon \nabla \Phi) &= {}-\textstyle\sum_{k} z_{k} e c_{k} ,\\
\eta \bm{\nabla^{2}} \vec{u} &= \nabla p + \textstyle\sum_{k} ( \kT \nabla c_{k} + z_{k} e c_{k} \nabla \Phi ) ,\\
\nabla \cdot \vec{u} &= 0 .
\end{aligned}
\end{equation}
Based on the same principles as Poisson-Boltzmann theory, this mean field model is valid for moderate concentrations of monovalent ions without permanent magnetic moments in aqueous solution at room temperature.~\cite{andelman95a, holm01a} It allows for net fluxes of ions and can be used in non-equilibrium situations. The Poisson-Boltzmann equation can be recovered as a special case of the stationary electrokinetic equations in situations where the boundary conditions imply a system in thermodynamic equilibrium.

\subsection{Geometry, Boundary conditions, System Parameters}

\begin{figure}[!h]
\centering
\includegraphics[scale=1.0]{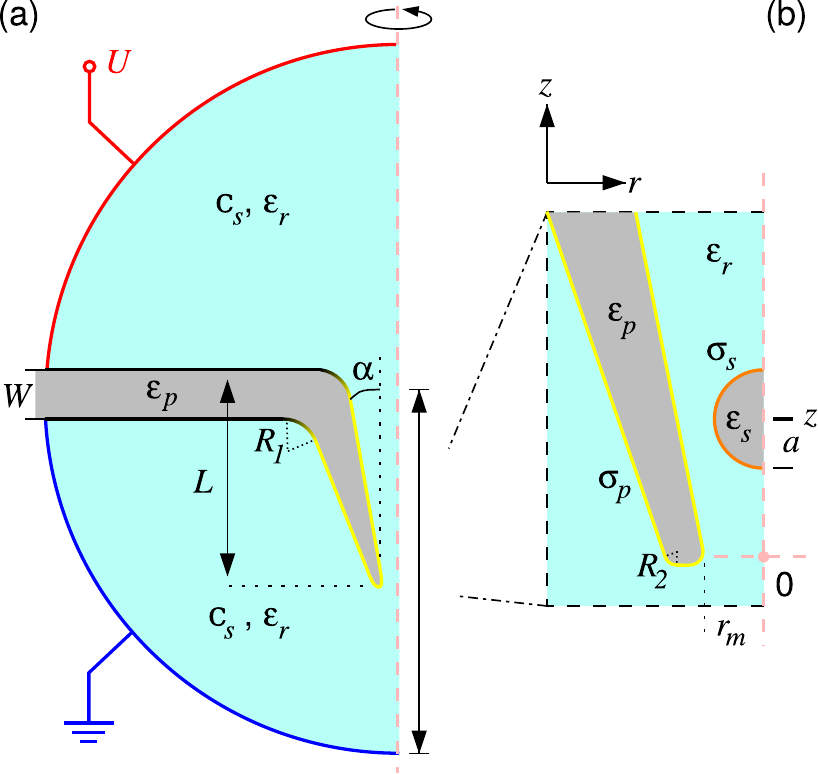}
\caption{Sketch of the conical nanopore geometry used in our numerical calculations. (a) The domain is rotationally symmetric around the red dashed axis. There are two hemispherical reservoirs (teal), separated by a barrier of radius $R$ and width $W$ (grey), from which a conical pore extrudes with length $L$ and opening angle $\alpha$. The pore and saline solution (bulk ionic concentration $c_s$) are dielectrics with relative permittivity $\varepsilon_\text{p}$ and $\varepsilon_\text{r}$, respectively. (b) Zoom-in on the pore orifice. The origin of the system is located at the narrowest part (the orifice), which has a radius $r_\text{m}$. The position of a spherical colloid with radius $a$ and relative permittivity $\varepsilon_\text{s}$ used for the investigations in \Cref{sec:translocation} is given by $z$. The surface charge of the pore $\sigma_p$ and the nanoparticle $\sigma_s$ are also indicated.}\label{fig:geometry}
\end{figure}

Let us now introduce the boundary conditions to the bulk equations of the previous paragraph that specify nanopore and nanoparticle. \Cref{fig:geometry} shows the system under investigation, a conical nanopore separating two reservoirs, which models the experimental setup of~\citet{Laohakunakorn15a}. We employ cylindrical coordinates: radial $r$ and axial $z$, with respective unit vectors $\hat{r}$ and $\hat{z}$. A conical nanocapillary with length $L = \SI{20}{\micro\m}$, angle $\alpha = \SI{5}{\degree}$, and orifice radii $r_{\text{m}}$ ranging from \SIrange{7.5}{150}{\nano\m}, extrudes from a circular barrier of radius $R = 2.5L$ and width $W = 0.15L$, separating two hemispherical reservoirs containing aqueous saline solution (teal), see Fig.\,\ref{fig:geometry}a. Circular arcs with a (smoothing) radius $R_1$ connect the capillary surface to the barrier (gray), see Fig.\,\ref{fig:geometry}a. Likewise, we use circular arcs with a smoothing radius $R_2$ to connect the inner and outer capillary surface to the flat front, see Fig.\,\ref{fig:geometry}. The relative permittivity of the solution is homogeneous and is assumed to be that of water, $\varepsilon_{\text{r}} = 78.5$, while the capillary and barrier have $\varepsilon_{\text{r}} = 4.2$ to match that of silica. The capillary carries a surface charge density $\sigma_{\text{p}} = \SI{-0.125}{e\,\nano\m^{-2}}$ (yellow), which agrees with typical experimental values.~\cite{Laohakunakorn15a, Laohakunakorn15b} In \Cref{sec:rectification}, we consider this nanocapillary without the nanoparticle present, focusing on the effect of physical parameters, such as, salt concentration and imposed electric field, as well as geometric parameters, such as the length $L$ and barrier width $W$.

In \Cref{sec:translocation} we place a spherical colloid of radius $a$, surface charge $\sigma_{\text{s}}$ (orange), and relative permittivity $\varepsilon_{\text{s}} = \varepsilon_{\text{p}} = 4.2$ close to the nanopore tip, see Fig.\,\ref{fig:geometry}b. The position of the spherical colloid (indicated using $z$, with $z=0$ the narrowest point of the nanopore) is constrained to the symmetry axis ($r=0$) and held fixed for each simulation. This effectively allows us to simulate the translocation of a nanoparticle through the conical nanopore by varying the position $z$ and computing the forces on the particle. The limitations of this approximation will be discussed further in \Cref{sec:translocation}, wherein we also study the ability of the conical nanopore to distinguish between various nanoparticles with different properties and under which conditions nanoparticle translocation can occur. 

For both of these systems, we impose no-slip boundary conditions ($\vec{u} = \vec{0}$) on the surface of the pore, barrier, and sphere; and vanishing normal stress at the edges of the reservoirs (red, blue), see Fig.\,\ref{fig:geometry}a. The condition of vanishing normal stress at the reservoir boundaries prevents momentum exchange with the reservoirs, while still allowing fluid flow into and out of the reservoirs. This is possible since momentum in Stokes' equations~\eqref{eq:stokes} is only transported through viscous stress and not through convection.

We include two ionic solute species with valencies $z_{\pm} = \pm 1$ and diffusion coefficients $D_{\pm} = \SI{2e-9}{\m^2s^{-1}}$, corresponding to those of K$^{+}$ and Cl$^{-}$. The ions' mobilities result from the Einstein-Smoluchowski relation~\cite{einstein1905a, smoluchowski1906a} with a temperature $T = \SI{298.15}{\kelvin}$. We apply a no-flux boundary conditions at the capillary, barrier, and sphere and impose bulk ionic concentrations $c_s$ ranging from \SIrange{1e-4}{1}{\Molar} at the reservoir boundaries, see Fig.\,\ref{fig:geometry}a.

We also use the reservoir boundaries to impose an external electric field, driving ion currents and EOF through the pore. We apply constant electric potential boundary conditions to both reservoir boundaries, setting the lower reservoir boundary (blue) to the reference voltage of $\SI{0}{\volt}$ and applying voltages in the range of \SIrange{-1}{1}{\volt} to the upper reservoir boundary (red). Consequently, positive voltages correspond to an electric field pointing from the capillary orifice into the bulk (in the direction of $-\hat{z}$), while for negative voltages the $E$-field points in the direction of $\hat{z}$. Henceforth, we will discuss the direction of the $E$-field in terms of the sign of the applied voltage. We employ surface-charge boundary conditions on the sphere (orange) and pore (yellow), while the barrier carries no surface charge (black) and there is a smooth transition between the capillary and the barrier, see Fig.\,\ref{fig:geometry}a.

\section{\label{sec:fem}Finite-Element Model}

The electrokinetic equations can be solved analytically in the limits of very high or very low salt concentration, yielding important results for the electrophoretic mobility of charged macromolecules and colloids under these conditions.~\cite{smoluchowski1906a, huckel1924kataphorese} For intermediate salt concentrations, analytic solutions are only tractable for a few highly symmetric systems and, even in those cases, often in the presence of only a single counterionic species. However, a wide variety of numerical methods capable of solving the electrokinetic equations in the general case exist.~\cite{capuani_discrete_2004, giupponi_colloid_2011, smiatek_mesoscopic_2011, schmitz_numerical_2012, medina_efficient_2015} In this section, we will introduce the algorithm used in this investigation, which is based on the finite-element method.

\subsection{Discretization}

To apply the FEM to the electrokinetic equations~\eqref{eq:summary}, one must first express this system of coupled non-linear partial differential equations (the strong formulation) in the so-called weak formulation. For the sake of brevity, we will demonstrate this procedure only for Poisson's equation~\eqref{eq:poisson}. In the weak formulation, one multiplies the terms in the equation by a test function $\varphi$ and integrates these expressions over the whole domain. A field $\Phi$ is considered a solution if it fulfills the resulting relation for any test function $\varphi$. In the case of Poisson's equation~\eqref{eq:poisson}, this relation reads:
\begin{align}
\label{eq:weak_form}
\int_\Omega \varphi \nabla^2 \Phi \, dV &= -\int_\Omega \varphi \varrho / \varepsilon \, dV \notag \\
\Leftrightarrow \int_\Omega \nabla \varphi \nabla \Phi \, dV &= \underbrace{\int_{\partial\Omega} \varphi \nabla \Phi \, d\vec{A}}_{= 0\text{, since } \varphi(\partial\Omega) = 0} + \int_\Omega \varphi \varrho / \varepsilon \, dV ,
\end{align}
assuming spatially homogeneous $\varepsilon$, requiring that the test function $\varphi$ vanishes at the domain boundary, and by using Green's first identity.

To numerically approximate a solution to the weak problem~\eqref{eq:weak_form}, one represents both the unknown electrostatic potential $\Phi(\vec{r})$ and the charge density $\varrho(\vec{r})$ in terms of a finite number of basis (ansatz) functions $b_i(\vec{r})$.
\begin{align}
\Phi(\vec r) = \textstyle\sum_k \Phi_k b_k(\vec r), \hspace{0.5cm} \varrho(\vec r) = \textstyle\sum_k \varrho_k b_k(\vec r) .
\end{align}

Also discretizing the test function $\varphi(\vec{r})$ using the same $b_i(\vec{r})$ allows one to test the relation~\eqref{eq:weak_form} with every basis function individually in the so-called Galerkin approach. Due to the linearity of relation~\eqref{eq:weak_form}, ensuring that it holds for all basis functions is equivalent to it being fulfilled for any test function $\varphi(\vec{r}) = \sum_i \varphi_i b_i(\vec{r})$. This leaves us with a system of equations of the form:
\begin{align}
\label{eq:poisson_fem}
 \resizebox{0.88\linewidth}{!}{$\underbrace{\sum_k \Phi_k \int_\Omega \nabla b_i(\vec r) \nabla b_k(\vec r) \, dV}_{\textstyle\sum_k \bar K_{ik} \Phi_k} = \underbrace{\sum_k \varrho_k / \varepsilon \int_\Omega b_i(\vec r) b_k(\vec r) \, dV}_{f_i(\varrho)} $} 
\end{align}

For computational efficiency, one typically ensures that matrix $\tens K$ is sparse by decomposing the domain into a mesh of small sub-domains and then using basis functions $b_i(\vec{r})$ that are only non-zero on one of these mesh elements. Furthermore, one typically uses polynomial ansatz functions on each of these sub-domains as this permits exact evaluation of the integrals.

Ultimately, the discretized problem consists of a system of linear equations for the coefficients of the solution $\Phi_k$ 
\begin{equation}
\label{eq:fem_discretised}
\tens K \vec \Phi = \vec f(\vec \varrho)
\end{equation}
where the coefficients $\Phi_k$ and $\varrho_k$ form the vectors $\vec \Phi$ and $\vec \varrho$, respectively, and the matrix $\tens K$ is given by~\eqref{eq:poisson_fem}.

\subsection{Coupling the different Equations}

Stokes' equation~\eqref{eq:stokes} and the diffusion-advection equations~\eqref{eq:concentrations} yield discretized equation systems similar to~\eqref{eq:fem_discretised}, the differences being that the operator $\tens K$ in the diffusion-advection equations depends on the solutions for the electrostatic potential and the flow velocity, and that the right-hand side of the discretized Stokes' equation is a non-linear function of the ionic concentrations and the electrostatic potential. The three coupled discretized equations can be combined into one fully-coupled system in the following way:
\begin{equation}
\label{eq:fem_fully_coupled}
\underbrace{\begin{bmatrix}
\tens K_1(\vec \Phi, \vec u) & 0 & 0 \\
0 & \tens K_2 & 0 \\
0 & 0 & \tens K_3
\end{bmatrix}
\begin{bmatrix}
\vec c \\
\vec \Phi \\
\vec u
\end{bmatrix}
-
\begin{bmatrix}
0 \\
\vec f_2(\vec c) \\
\vec f_3(\vec \Phi, \vec c) \\
\end{bmatrix}}_{\vec{F}([\vec c, \vec \Phi, \vec u])}
=
\begin{bmatrix}
0 \\
0 \\
0 \\
\end{bmatrix} .
\end{equation}
Here the first row comprising $\tens K_1$ represents the two discretized diffusion-advection equations~\cref{eq:concentrations}, the second row comprising $\tens K_2$ and $\vec{f}_2$ consists of the discretized Poisson's equation as derived in~\cref{eq:fem_discretised}, and the third row comprising $\tens K_3$ and $\vec{f}_3$ represents Stokes' equations~\cref{eq:stokes}. The discrete vector solved for consists of $\vec{c}$ containing the ionic concentrations coefficients in the FEM approximation, while the coefficients for the electrostatic potential are contained in $\vec{\Phi}$, and the coefficients for the flow velocity and the hydrostatic pressure are contained in $\vec{u}$.

We use a combination of third order polynomial ansatz functions for the ionic concentrations, second order for the electrostatic potential, second order for the hydrostatic pressure, and third order for the fluid flow. Ideally, one would use an implementation of the finite-element method capable of adaptively choosing the degree of the ansatz functions for every mesh element. This adaptivity leads to significant improvements in the efficiency over methods with fixed ansatz functions.~\cite{guo_h-p_1986} Unfortunately, such an implementation was not available to the authors at the time of this investigation.

\subsection{Solving the coupled, non-linear System}

As discussed in the previous section, discretizing the electrokinetic equations~\eqref{eq:summary} using the FEM leads to a system of non-linear, coupled equations whose shape is given in~\eqref{eq:fem_fully_coupled}. This system can be solved using a method for non-linear root-finding, such as Newton's method. Applying Newton's method to the discrete system~\eqref{eq:fem_fully_coupled} yields the iteration scheme:
\begin{equation}
\label{eq:fem_newton}
\vec D\vec F \left(
\begin{bmatrix}
\vec c_n \\
\vec \Phi_n \\
\vec u_n
\end{bmatrix}
\right)
\begin{bmatrix}
\vec c_{n+1} - \vec c_n \\
\vec \Phi_{n+1} - \vec \Phi_n \\
\vec u_{n+1} - \vec u_n
\end{bmatrix}
=
-\vec F \left(
\begin{bmatrix}
\vec c_n \\
\vec \Phi_n \\
\vec u_n
\end{bmatrix}
\right) ,
\end{equation}
with a slightly unwieldy, but nevertheless explicitly known expression for the total derivative $\vec D \vec F$ of the operator $\vec F$ as defined in~\eqref{eq:fem_fully_coupled}.

Whether the iteration scheme~\eqref{eq:fem_newton} converges depends on the prescribed boundary conditions, the mesh, and the initial guess for the solution. For the system introduced in \Cref{sec:ekin_eq} with a mesh suitably discretizing the double layers, and negative applied voltages, convergence is achieved with a very simple initial guess for the solution, consisting of the constant fields $\vec u = 0$, $\Phi = 0$, and $c_\pm = c_s$. Positive applied voltages require a better initial guess. We start with simulations for negative applied voltages as low as $U_\text{min} = \SI{-1}{\volt}$, increasing the voltage in steps of $\Delta U = \SI{0.1}{\volt}$ up to $U_\text{max} = \SI{1}{\volt}$, using the each run's solution as the initial guess for the next.

There is a number of powerful and user-friendly simulation codes based on the finite-element method to carry out these simulations; we use the FEM simulation suite COMSOL Multiphysics 5.1.

\subsection{The Mesh}

\begin{figure}[!htb]
\includegraphics[width=\columnwidth]{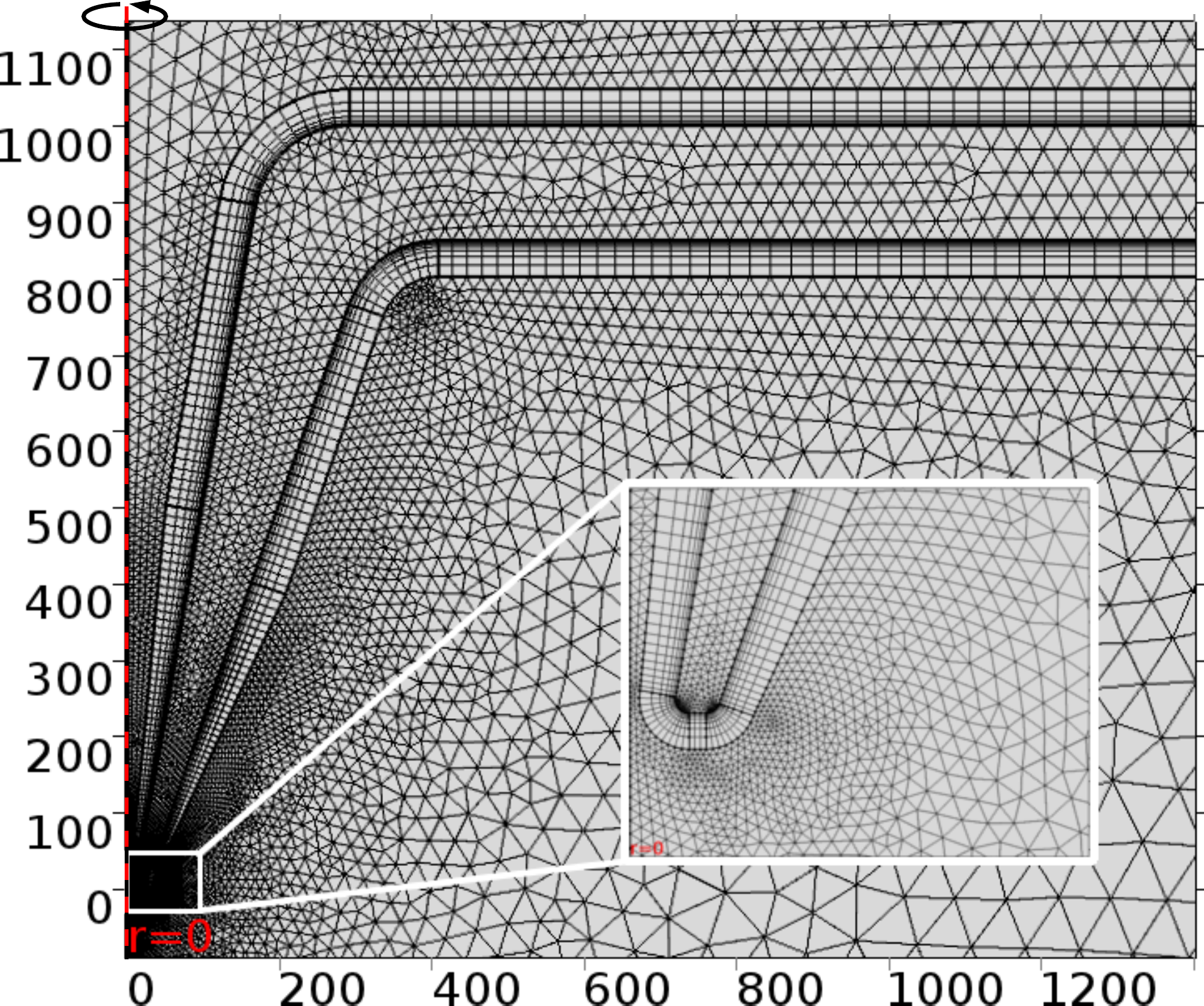}
\caption{Mesh used in the discretization of the nanocapillary geometry. The mesh consists of a layer with a thickness of 5 Debye lengths at the capillary and barrier surface consisting of quad elements. The size of these elements in the normal direction to the surface is chosen such that the first Debye length contains 5 elements and the element size increases exponentially just like the electric potential decreases exponentially. The remaining volume is discretized using triangular elements. The inset shows a zoom-in of the region directly at the capillary tip. The dashed red line indicates the rotational symmetry axis for this axisymmetric problem.}
\label{fig:mesh}
\end{figure}

As we will show in \Cref{sec:rectification}, the length of the pore needs to be $L = \SI{20}{\micro\m}$ to reduce finite-size effects to acceptable levels. The physics of the problem require us to model the dynamics on the whole domain of diameter $2R = \SI{100}{\micro\m}$, while simultaneously resolving the nanometer-sized double layers. Achieving this requires a carefully crafted mesh as depicted in Fig.\,\ref{fig:mesh}. Highly anisotropic triangular elements lead to ill-conditioned equation systems after discretization through the FEM, which is why we discretize the double layers using quad elements, as we found this approach useful for other finite-element calculations as well.~\cite{de_graaf_diffusiophoretic_2015, rempfer_limiting_2016, kreissl16, rempfer_selective_2016, brown16} This allows us to discretize the large gradients in the normal direction of the charged surfaces optimally, while at the same time taking advantage of the slow variation of the fields parallel to these surfaces. The thickness of the quad layer linearly decreases towards the capillary orifice, since this layer would otherwise make contact with the simulation domain boundary at $r = 0$, which would introduce at least one highly anisotropic triangular element.

There are algorithms for adaptive meshing, which automatically refine the grid during the iterative solution process based on the solution's gradients. In our experience, these algorithms perform well for simpler systems, such as the electrophoresis of a charged sphere of a size comparable with the Debye length. Unfortunately, the implementation of these algorithms in COMSOL is only compatible with triangular meshes, which leads to excessive numbers of mesh elements for this system.

\section{\label{sec:applications}Rectification and Transport through Nanopores}

\begin{figure*}[htb!]
\begin{minipage}[t]{\textwidth}
\includegraphics[width=8.64468cm]{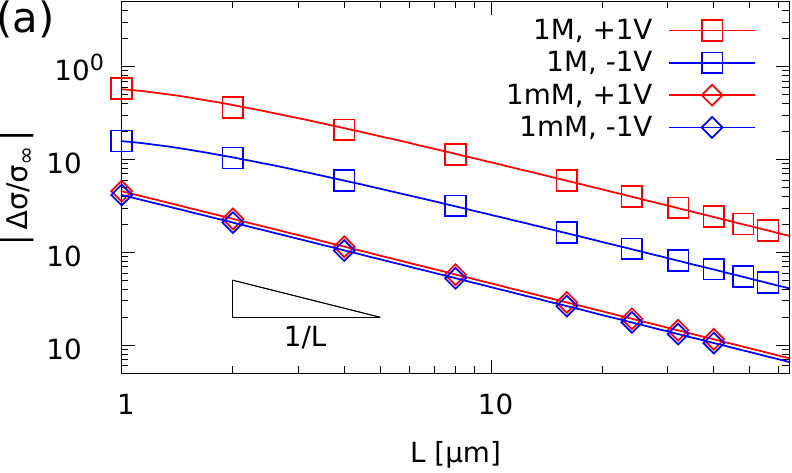}\hfill\includegraphics[width=8.64468cm]{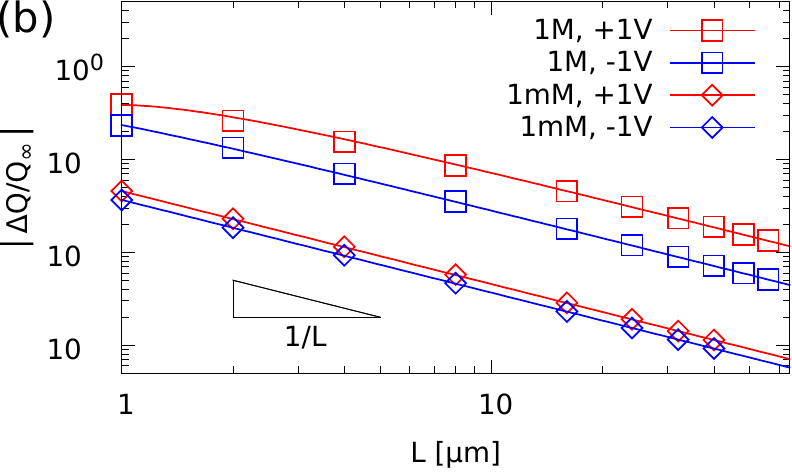}
\caption{\label{fig:finite-size}Finite-size scaling of the electric conductivity (a) and electro-osmotic flow (b) through the nanocapillary. For both quantities, the relative deviation to the values for an infinitely long capillary scales as the inverse length of the capillary. Higher-order deviations are only noticeable for capillaries shorter than \SI{2}{\micro\m} at high salt concentrations. For a length of \SI{20}{\micro\m}, the maximum relative error is limited to \SI{4}{\percent} for positive voltages at high salt concentrations; for other combinations of applied voltage and salt concentration it is much lower.}
\end{minipage}
\end{figure*}

In this section, we consider two specific physical problems, rectification and translocation through a nanopore, to which we can apply the electrokinetic equations and finite element method introduced above. We start with current and flow rectification in a conical nanopore, to demonstrate the quality of our methods, before switching to the more complicated translocation problem.

\subsection{\label{sec:rectification}Rectification Effects in Glass Nanocapillaries}

Before investigating the more interesting problem of colloidal translocation and the possibility to use this nanocapillary as a device to characterize nanoparticles, we characterize the empty capillary system in terms of its conductivity and electro-osmotic flow properties. This type of nanocapillary system has been subject to extensive experimental study by Keyser \emph{et al.}~\cite{laohakunakorn_landausquire_2013, laohakunakorn_electroosmotic_2015-1} That is, we specifically tailored our geometry to be representative of their experimental system.

We have performed a detailed analysis of influence of geometric parameters, including the smoothing of the nanopore tip and capillary-barrier transition, the barrier thickness, the shape of the electrodes, the surface charge smoothing, and the reservoir size. For a length of $L = \SI{20}{\micro\m}$ there is a \SI{4}{\percent} deviation from the result for an infinitely long pore, as the finite-size-scaling results in Fig.\,\ref{fig:finite-size} show. The relative error for both, the electric current, and the electro-osmotic flow is the highest for positive voltages (\SI{1}{\volt}) at high salt concentrations (\SI{1}{\Molar}). For other combinations of applied voltage and salt concentration, the errors are significantly reduced. Based on these scaling results, we carry out all further simulations with pores of length $L = \SI{20}{\micro\m}$. The barrier thickness has no significant influence on the current and flow in the system.~\cite{Ehrhardt03} At the previously established length of \SI{20}{\micro\m}, the influence of details of the capillary's back entrance is smaller than the finite-size errors,~\emph{i.e.}, the part of the capillary closest to the upper reservoir. These details include the corner smoothing radius and whether the capillaries surface charge transitions into the uncharged barrier smoothly or discontinuously, see Fig.\,\ref{fig:geometry}. While the reservoir size does influence simulation results significantly, these effects can be eliminated without incurring much additional computational cost by extending the reservoir radius to $R = 2.5L = \SI{50}{\micro\m}$ with a very coarse mesh. The details of the nanopore tip lead to more significant variation of the current and flow, as is to be expected. However, we found that the variation of the measured currents and flow velocities is less than \SI{2}{\percent} for tip smoothing radii $R_2$ in the range of \SIrange{1}{4}{\nano\m}, which is why we use $R_2 = \SI{3}{\nano\m}$ in all following simulations.

Another question concerns the importance of the advective transport of ionic species in the diffusion-advection equations~\eqref{eq:concentrations}. It is known for nano-scale systems that the diffusive transport dominates the ion dynamics. Neglecting the advective contribution to the ion flux reduces the computational cost of these simulations significantly as it allows the diffusion-advection and the electrostatics equations~\eqref{eq:poisson} to be solved separately from Stokes' equation~\eqref{eq:stokes}.

\begin{figure*}
\begin{minipage}[t]{\textwidth}
\includegraphics[width=8.64468cm]{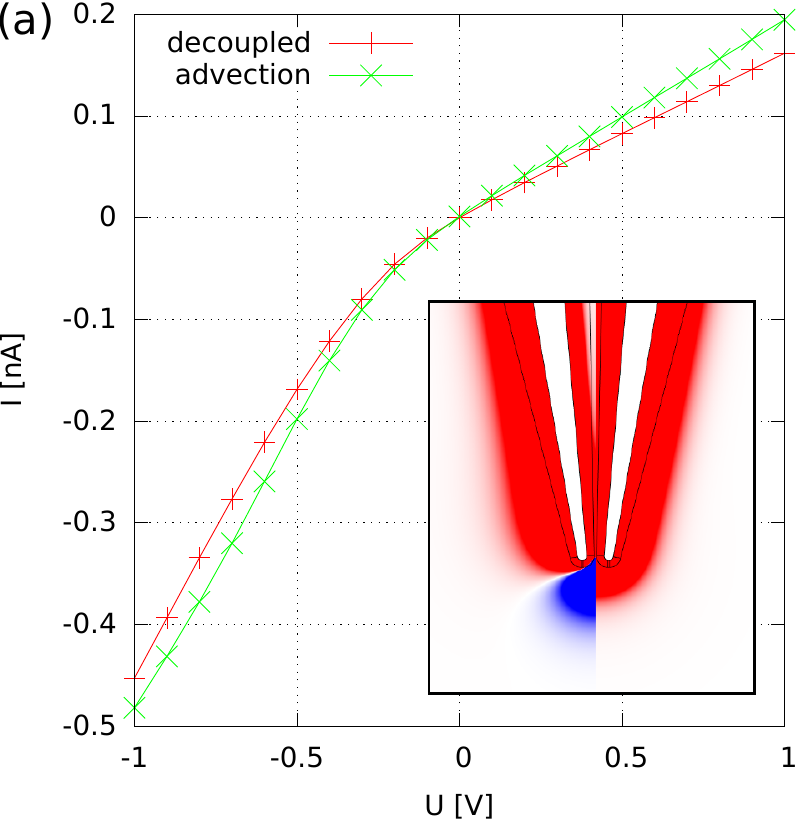}\hfill\includegraphics[width=8.64468cm]{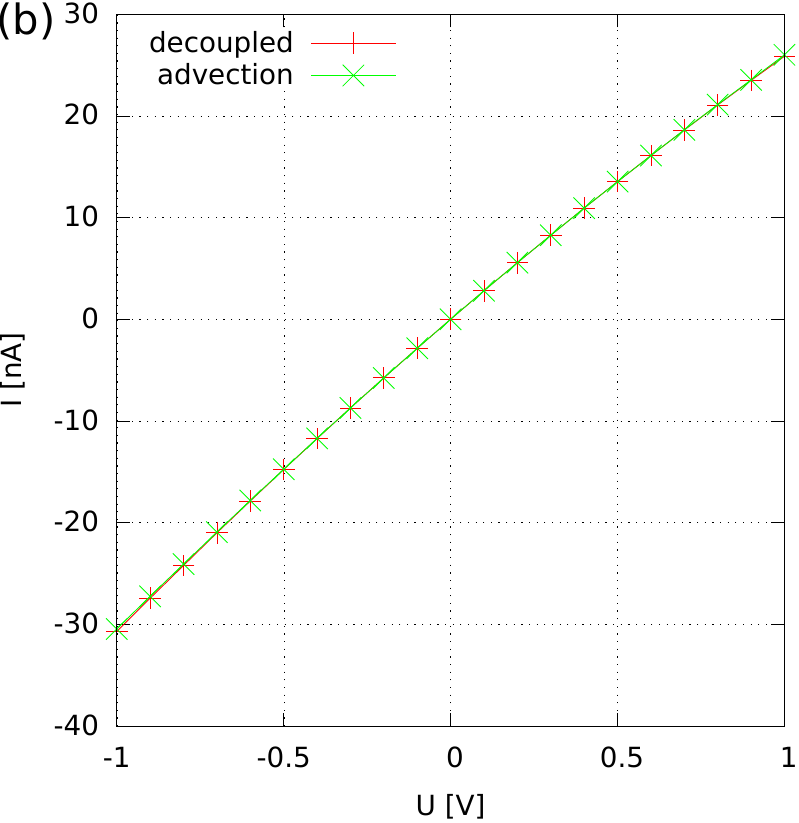}
\caption{\label{fig:advection}Net electric current conducted through the nanocapillary at a low salt concentration of \SI{1}{\milli\Molar} (a) and at a high salt concentration of \SI{1}{\Molar} (b). The inset in (a) shows the polarization effect of the net charge density (neglecting advection). Its left half depicts the (local) net charge density for a negative voltage of \SI{-1}{\volt}, while the right half depicts the charge density for a positive voltage of \SI{1}{\volt}.}
\end{minipage}
\end{figure*}

To verify whether this simplification is valid in this much larger system, we run simulations at low (\SI{1}{\milli\Molar}) and high (\SI{1}{\Molar}) salt concentrations with and without the advective contribution to the ionic flux and measure the ionic current through the pore. We expect the largest deviations to occur at low salt concentrations and with high voltages. The electro-osmotic flow velocities for low salt concentrations exceed those for high salt concentrations. In addition to that, a lower salt concentration leads to a larger Debye length, which places more ions further from the capillary surface, where the flow velocities are higher. Both these effects lead to a stronger influence of the flow on the ionic distributions. \Cref{fig:advection}a demonstrates that the errors are indeed the largest for low salt concentrations and high voltages, but even for these parameters, the errors do not exceed the ones from the finite-size effect. That is why we neglect the advective ion flux in all further simulations.

\Cref{fig:advection} also shows the pronounced rectification effect for the ionic current exhibited by conical glass nanocapillaries. This effect is due to polarization at the tapered capillary tip.~\cite{Laohakunakorn15b} The inset of Fig.\,\ref{fig:advection}a depicts the net charge of both ionic species for negative applied voltage (left) and positive applied voltage (right). Red signifies positive, while blue signifies negative excess local charge. Not shown here is the net density of charge carriers in the capillary tip, which modulates the conductivity of the capillary and differs strongly for opposite voltages. With increasing salt concentration, the double layers shrink and their relative contribution to the conductivity decrease. When the Debye length is small, the electrolyte in the capillary is bulk-like and no asymmetry in the direction of the applied electric field is possible. \Cref{fig:advection}b demonstrates that the rectification effect is indeed significantly reduced at a higher salt concentration of \SI{1}{\Molar}. 

These rectification effects have previously been studied in similar geometries by Keyser~\textit{et al.}~\cite{laohakunakorn_electroosmotic_2015-1, laohakunakorn_electroosmotic_2015} and White~\textit{et al.}~\cite{white_ion_2008} by theory and simulation. Our results qualitatively agree with these previous studies, but there are quantitative differences stemming from the different system geometries: White~\textit{et al.} work with solid state nanopores, while Keyser~\textit{et al.}\ use the same capillary geometry as we do, but they do not model the upper reservoir, instead imposing the boundary condition some distance up in the capillary. Since we have found a strong influence of the size of this reservoir on our simulations, the latter may induce strong finite-size effects. 

In our final investigation of the empty pore system, we quantify the rectification ratio of the ionic current and the electro-osmotic flow as a function of the salt concentration (given in terms of the Debye length). The rectification ratio is the absolute of the ratio of the current (or flow) for $U = \SI{1}{\volt}$ and $U = \SI{-1}{\volt}$. That is, the current rectification is defined as $R_\text{I} = | I(\SI{-1}{\volt}) / I(\SI{1}{\volt}) |$ with the net current $I$. Similarly the flow rectification is $R_\text{Q} = | Q(\SI{1}{\volt}) / Q(\SI{-1}{\volt}) |$ with the net water flow $Q$ (measured through the narrowest part of the tip). The ratio of the current is inverted, so that both rectification ratios lie in the interval $[1,\infty)$. We expect the rectification ratio for the ionic current $R_\text{I}$ to approach 1 (no rectification) for large and small Debye lengths, where large and small is relative to the features of the geometry. 

\begin{figure*}
\begin{minipage}[t]{\textwidth}
\includegraphics[width=8.64468cm]{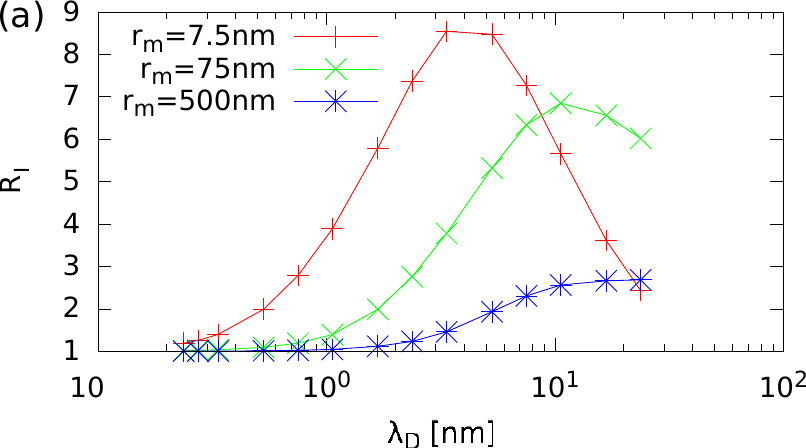}\hfill\includegraphics[width=8.64468cm]{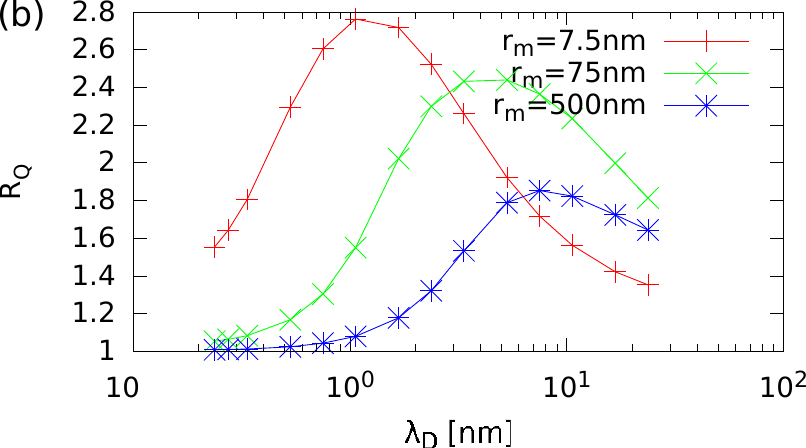}
\caption{\label{fig:rectification}Rectification ratio of the ionic current $R_\text{I}$ (a) and the electro-osmotic flow velocity $R_\text{Q}$ (b), both measured at the nanocapillary orifice. The rectification ratios are given as functions of the Debye length $\lD$ and the orifice diameter $r_\text{m}$. The rectification ratios asymptotically approach 1 (no rectification) in the limits of infinitely small (Smoluchowski) and infinitely large Debye length (H\"uckel). Significant rectification ratios are achieved in the intermediate regime, where the Debye length is comparable to the orifice diameter.}
\end{minipage}
\end{figure*}

\Cref{fig:rectification} shows these rectification ratios for capillaries with three different orifice diameters: $r_\text{m} = \SI{7.5}{\nano\m}$, $r_\text{m} = \SI{75}{\nano\m}$, and $R_\text{I} = \SI{500}{\nano\m}$. \Cref{fig:rectification}a depicts the electric current's rectification ratio $R_\text{I}$, while Fig.\,\ref{fig:rectification}b depicts the rectification ratio of the EOF $R_\text{Q}$. We now explain the observed peaks and structures in the rectification ratios.

In the limit of infinitely small Debye length, the Smoluchowski limit of high salt concentration,~\cite{smoluchowski1906a} the ionic concentrations assume their bulk values everywhere, independently of the applied voltage. The conductivity then does not depend on the direction of the applied voltage. Electro-osmotic flow is created in the double layer. The double layer is the only part of the volume, where the hydrodynamic driving forces of the two ionic species do not cancel. When the double layer is thin compared to the surface geometry features, the situation matches that of a flat wall for which there is no asymmetry.

In the limit of infinitely large Debye length, the so-called H\"uckel limit,~\cite{huckel1924kataphorese} there are only counterions and their distribution extends infinitely far from the charged object (the capillary). In the vicinity of the capillary and on the scale of its size, the ion density is constant and independent of the applied voltage, leading to no rectification. Note that we do not reach this limit in all cases, as the extreme differences in length scale that are obtained for such low salt concentrations make the electrokinetic equations difficult to solve in the nanopore geometry. 

In the intermediate regime for the Debye length, we expect the conical shape of the nanopore to break the symmetry, which creates rectification ratios $R_\text{I,Q} \ne 1$ for the electric current and fluid flow. As expected, the maximum lies at higher values for $\lD$, which is comparable to the capillary orifice diameter --- the relevant geometric scale. This optimum in rectification could prove relevant to tuning the sensitivity of this nanopore system for particle characterization, since, as we will see next, translocation signals are strongest for high salt concencentrations (or equivalently small $\lD$).

\subsection{\label{sec:translocation}Colloid Translocation}

After having validated our simulation model with the empty nanocapillary, we can move on to a more challenging problem. In this section, we place a spherical colloid along the symmetry axis (see Fig.\,\ref{fig:geometry}b) and systematically investigate the influence of salt concentration, applied voltage, particle surface charge, and particle size on the translocation of the colloid and on the ionic current. We consider only negative surface charges for the capillary, as is typical in experiments,~\cite{lan_nanoparticle_2011, german_controlling_2013, lan_effect_2014, Laohakunakorn15a, Laohakunakorn15b} and a like-charged nanosphere to prevent the particle sticking to the pore wall.

We do not directly simulate the dynamics of the sphere translocation. Instead, we utilize the separation of time scales between the development of the concentration profiles and fluid flow on the one hand, and the motion of the sphere on the other hand to decouple these two problems. By measuring the force at each point along the $z$-axis, the presence of force traps and barriers can be detected. Note that we ignore thermal fluctuations, which could move the sphere away from the symmetry axis. This reduction is necessary since solving the full three-dimensional (3D) problem (as for the low-aspect-ratio pores of~\citet{Tsutsui16}) is not computationally feasible.

\begin{figure}[!h]
\centering
  \includegraphics[scale=1.0]{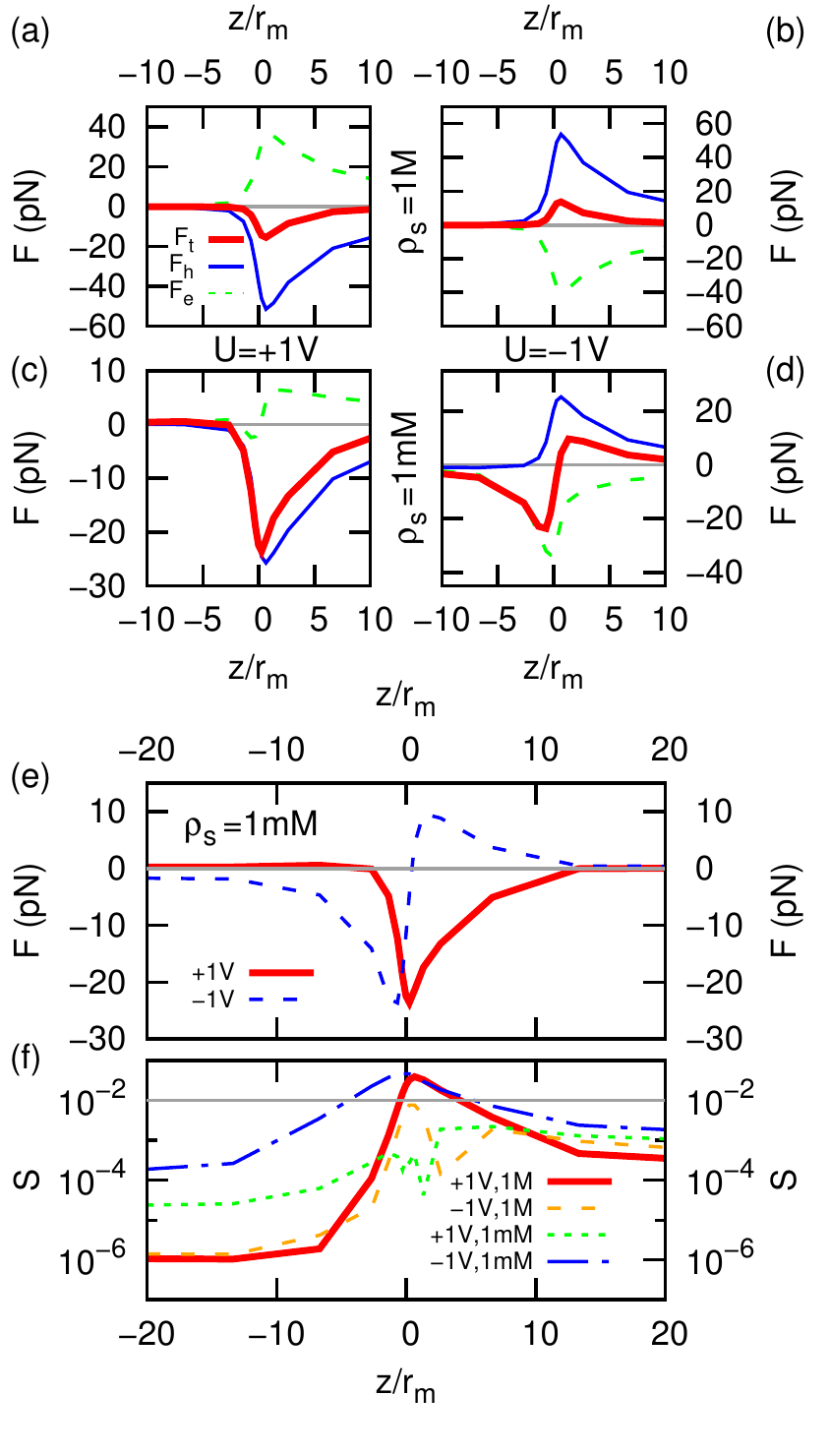}
  \caption{\label{fig:translocation}The force $\vec F$ experienced by a sphere of radius $a=\SI{3.5}{\nano\m}$ with surface charge $\sigma_\text{s} = \SI{-0.125}{e\,\nano\m^{-2}}$, and the current signal strength $S$ measured during the translocation event as a function of the sphere position $z$ divided by the pore radius $r_\text{m} = \SI{7.5}{\nano\m}$. (a-d) The total $\vec{F}_{\text{t}}$ (red, thick solid), hydrodynamic $\vec{F}_{\text{h}}$ (blue, thin solid), and electric $\vec{F}_{\text{e}}$ (green, thin dashed) force in pN. The panels show: (a) a positive applied voltage $U = \SI{+1}{\volt}$ and a bulk salt concentration of $c_s = \SI{1}{\Molar}$, (b) $U = \SI{-1}{\volt}$ and $c_s = \SI{1}{\Molar}$, (c) $U = \SI{+1}{\volt}$ and $c_s = \SI{1}{\milli\Molar}$, and (d) $U = \SI{-1}{\volt}$ and $c_s = \SI{1}{\milli\Molar}$; as also indicated using the labels. (e) Close-up of the total force $\vec{F}_{\text{t}}$ for $c_s = \SI{1}{\milli\Molar}$; $U = \SI{+1}{\volt}$ (red solid) and $U = \SI{-1}{\volt}$ (blue dashed). (f) The current signal strength $S$ during translocation for $U = \SI{+1}{\volt}$ and $c_s = \SI{1}{\Molar}$ (red), $U = \SI{-1}{\volt}$ and $c_s = \SI{1}{\Molar}$ (orange), $U = \SI{+1}{\volt}$ and $c_s = \SI{1}{\milli\Molar}$ (green), $U = \SI{-1}{\volt}$ and $c_s = \SI{1}{\milli\Molar}$ (blue); the horizontal grey line indicates \SI{1}{\percent}.}
\end{figure}

In order to determine the effect of the sphere moving through the pore on the symmetry axis, we compute the total force on the sphere $\vec{F}_{\text{t}}$, which we can split into an electric $\vec{F}_{\text{e}}$ and a hydrodynamic (mechanical) $\vec{F}_{\text{h}}$ component. The force is positive if it points in the $\hat{z}$ direction (from the lower reservoir into the pore orifice). We also consider the current $I$ through the pore (as a function of the sphere position) and the signal strength $S = \vert \left( I - I_{0} \right) / I_{0} \vert$, with $I_{0}$ being the current in the absence of the sphere. We also define $S_{m} \equiv \max_{z} S$, the maximum signal strength. \Cref{fig:translocation} shows the force and signal strength $S$ curves for four combinations of salt concentration and applied potential: $c_s = \SI{1}{\Molar}$ and $\SI{1}{\milli\Molar}$, and $U = \pm \SI{1}{\volt}$. In all cases the colloid radius is $a = \SI{3.5}{\nano\m}$, and the surface charge density is $\sigma_{\text{s}} = \SI{-0.125}{e\,nm^{-2}}$; the sphere carries the same surface charge density as the pore and is roughly half the orifice diameter in size. We discuss these results next.

At high salt concentration, the hydrodynamic force coming from the EOF always dominates the electric force, and the particle moves along the direction of the applied field (since the flow is in that direction). For $U = \SI{+1}{\volt}$, the movement is from the top reservoir into the bottom one and vice versa for $U = \SI{-1}{\volt}$, see Fig.\,\ref{fig:translocation}b. This means that the sphere can translocate through the nanopore at high salt concentrations. In all cases there is a clear current signal jump as the particle comes close to the pore orifice. This indicates that the region near the orifice contributes most to the signal and not the main body of the pore, even at the relatively small pore angle of $\alpha = \SI{5}{\degree}$. This is in agreement with the findings of \citet{Tsutsui16} for short nanopores. 

We obtain a signal strength of $S_{m} = \num{0.040}$ for $U = \SI{+1}{\volt}$ and $S_{m} = \num{0.008}$ for $U = \SI{-1}{\volt}$ for the $c_s = \SI{1}{\Molar}$ case, see Fig.\,\ref{fig:translocation}; the former providing a clearer measurement. The base current $I_{0}$ is comparable in both cases: $I_{0} = \SI{26}{\nano\ampere}$ ($U = \SI{+1}{\volt}$) and $I_{0} = \SI{-31}{\nano\ampere}$ ($U = \SI{-1}{\volt}$), respectively. Hence, the difference in $S_{m}$ is due to the current carrying species. The presence of the negatively charged sphere in the pore orifice effectively doubles the amount of double layer, which has overall more K$^{+}$ ions than Cl$^{-}$ ions, with the K$^{+}$ ions being closer to the surface. Thus, when the sphere is in the pore orifice, a new rectifying structure is formed, which has the same current rectification properties as the original, hence the observed difference in $S_{m}$. 

At low ionic strength, the particle displays interesting behavior, see Fig.\,\ref{fig:translocation}d/e. When a positive voltage is applied, the total force in both reservoirs points towards the orifice, as can be seen from the change in sign of the force at $z \approx \SI{-20}{\nano\m}$ ($z/r_\text{m} \approx 3$). This means that the particle becomes trapped at this point. The change in sign of the total force can be explained as follows. In the lower reservoir, the electric force dominates over the hydrodynamic force close to the orifice, whereas inside the pore the hydrodynamic force dominates (close to the orifice); since both forces point in opposite directions, the net effect is the formation of a trapping zone. The reason why the direct electric force is greater than the hydrodynamic force coming from the EOF in the lower reservoir is that the double layer extends outward from the pore at low ionicity.~\cite{Laohakunakorn15b} Our trapping is thus different in nature from that observed when an external hydrostatic pressure difference is applied.~\cite{Luo14,german_controlling_2013,lan2011pressure}

Further investigation is required to determine whether the trapping point is an inflection point or a true trap that extends beyond the $z$-axis in the radial direction; this requires fully resolved 3D simulations which go beyond the scope of the current investigation. However, this feature is problematic in either case, as the system becomes unsuited to extract particle properties by means of translocation. For pores of a significantly bigger radius, one can model the translocating particle implicitly by giving an equation of motion in terms of the electric field and flow field from the FEM simulation. This approach can be applied to off-axis translocation without added computational cost.~\citep{rempfer_selective_2016}

For a negative applied voltage at low ionic strength, a large current signal is observed, see Fig.\,\ref{fig:translocation}. Unfortunately, in this case the force points away from the pore orifice in both reservoirs, and the inversion in direction takes place at $z = \SI{0}{\nano\m}$, see Fig.\,\ref{fig:translocation}e. Thus, the particle is repelled from the pore orifice in both reservoirs, preventing translocation. The reason for this is the same as for the positive applied voltage at low ionicity, however, the direction of the effect is now reversed. The observation of a bi-directional repulsion leads us to conclude that translocation is not possible in this case. Thus, to obtain translocation and a decent signal, high salt concentrations and positive voltages are required (for negatively charged pores and particles).

\begin{figure}[!h]
\centering
  \includegraphics[scale=1.0]{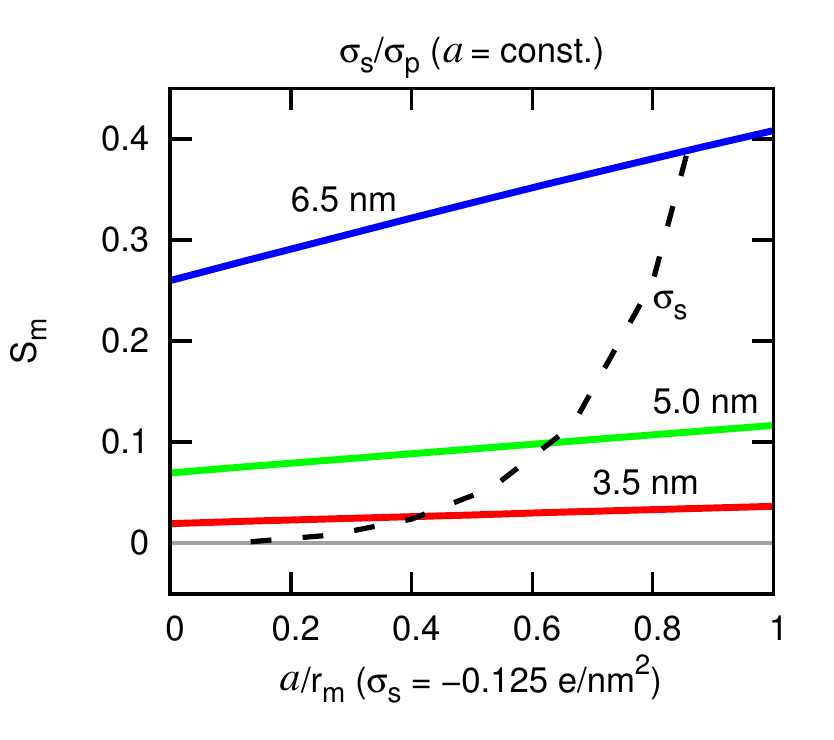}
  \caption{\label{fig:signal}The maximum signal strength $S_{m}$ as a function of the sphere size $a$ (black dashed) and the sphere charge $\sigma_\text{s}$ (solid lines) for a system with $U = \SI{+1}{\volt}$ and $c_s = \SI{1}{\Molar}$. The surface charge is fixed to \SI{-0.125}{e\,\nano\m^{-2}} when varying the sphere radius. The radius is kept fixed to $a = \SI{3.5}{\nano\m}$ (red), $a = \SI{5.0}{\nano\m}$ (green), and $a = \SI{6.5}{\nano\m}$ (blue) when varying the surface charge.}
\end{figure}

Considering the above, we study the maximum signal strength $S_{m}$ as a function of the sphere radius $a$ and surface charge $\sigma_{\text{s}}$ for $U = \SI{+1}{\volt}$ and $c_s = \SI{1}{\Molar}$, see Fig.\,\ref{fig:signal}. For a given $\sigma_{\text{s}} = \SI{-0.125}{e\,\nano\m^{-2}}$, we observe that the signal strength $S$ displays a nearly cubic dependence on the sphere radius $a$ ($S_{m} \propto a^{3}$), with $S_{m}$ approaching \num{0.4} for a sphere that only has a \SI{1}{\nano\m} pore clearance at the orifice. However, the dependence of $S_{m}$ on the surface charge is linear ($S_{m} \propto \sigma_{\text{s}}$), as can be seen in Fig.\,\ref{fig:signal}, and relatively small. Indeed, for the entire range of surface charges we investigated, the ranges in $S_{m}$ for sphere sizes that are \SI{1.5}{\nano\m} apart are completely separate. This demonstrates that one can use the conical nanopore as a characterization device that can quantify the radius of nanoparticles by measuring the current signal to within a precision of at least \SI{1}{\nano\m}. However, a separate characterization of the charge polydispersity could in principle reduce this number. The above observations are in line with the work by Lan~\textit{et al.},~\cite{lan_nanoparticle_2011} in which a similar dependence of the current signal on the particle size was considered. Lan~\textit{et al.} also considered influence of the particle size on the length of the translocation effect. However, we do not do so here, as our quasi-static method prevents us from establishing the translocation time with sufficient accuracy.

\section{\label{sec:conclusion}Conclusions}

Summarizing, we have presented a simulation model using the finite-element method to solve the electrokinetic equations in a geometry representing the tapered glass-nanocapillaries (conical nanopore), that are used \emph{e.g.}\ by the Keyser group.~\cite{thacker_studying_2012, li_single_2013} We validated this model by reproducing the known rectification properties for the ionic current and electro-osmotic flow through these nano-capillaries. We then used this model to investigate the translocation of a (negative) nanoparticle through a negatively charged conical nanopore. 

In the absence of nanoparticles, the nanopore functions as a current and flow rectification device, when placed in a saline solution and when and external electric field is applied over it. We reproduce the rectification results originally observed in Refs.~\cite{laohakunakorn_landausquire_2013, laohakunakorn_electroosmotic_2015} and expand on these. This expansion is made possible in large part by our improved meshing and modification of fluid forcing term in Stokes' equation, which strongly reduces spurious flow.~\cite{rempfer_limiting_2016} The finite-size effect on the flow and current are studied and found to asymptotically scale with the inverse length of the capillary. To faithfully represent the experimental systems with capillary lengths in the \si{\milli\m} to \si{\centi\m} range requires simulation of a \SI{20}{\micro\m} capillary and \SI{100}{\micro\m} of the surrounding reservoir. Our reduced spurious flow algorithm also allowed us demonstrate that advection of the ions can be safely ignored. Finally, the optimum ionicity for achieving rectification is identified and the various limiting behaviors are discussed.

Next we considered nanoparticle translocation through our conical nanopore geometry, without an externally applied pressure difference. Here we find similar effects as originally observed (both experimentally and numerically) in Refs.~\cite{lan_nanoparticle_2011, li_single_2013, lan_effect_2014}. Varying the environmental parameters, such as ionicity and applied voltage (or equivalently direction of the externally applied $E$-field), has allowed us to identify three key features of this system, which are of significant interest to experiments on nanopore translocation.

First, translocation can only take place at high ionic strength. For low ionic strength we observe both particle trapping at the pore orifice and particle repulsion from the orifice, depending on the direction of the applied electric field. It might be possible to overcome this effect by applying a hydrostatic pressure difference between the two reservoirs.~\cite{Luo14, german_controlling_2013, lan2011pressure} Second, the signal strength --- defined as the change in electric current through the pore during the translocation, with respect to the base current without the sphere --- is largest when a positive voltage ($E$-field pointing out of the pore orifice into the bulk) is applied at high ionic strength or a negative voltage is applied at low ionic strength. Since the latter is excluded for translocation, one can only use the nanopore to perform particle characterization by translocation at high ionicity and positive voltage. Third, the signal strength is sufficient to observe the particle translocation and has a cubic dependence on the radius of the particle, while it is only weakly linearly dependent on the surface charge of the particle. This result is similar to that obtained by Lan~\emph{et al.}~\cite{lan_nanoparticle_2011} The signal strength can be as much as \SI{40}{\percent} of the base current for large particles. Therefore, our calculations indicate that the conical nanopore is suited as a particle characterization device, which can sensitively discriminate between particles with different radii. 

In conclusion, we have demonstrated the versatility of finite-element calculations to study electrokinetic phenomena for the conical nanopore geometry. Our numerical work shows that the long conical nanopores produced in the Keyser group~\cite{thacker_studying_2012, li_single_2013, laohakunakorn_landausquire_2013, laohakunakorn_electroosmotic_2015} are suited for nanoparticle characterization. They should give similar current signals as observed for other nanoporous systems,~\cite{lan_nanoparticle_2011, lan_effect_2014, german_controlling_2013} provided the optimized parameters identified here are employed. Future work will focus on extending our calculations to study the observed trapping ability of the pore, as the precise localization of particles is also of significant benefit to characterization studies.

\section*{Acknowledgments}

GR and CH thank the DFG for funding through the SFB716/TPC.5. JdG gratefully acknowledges support from an NWO Rubicon Grant (\#680501210) and a Marie Sk{\l}odowska-Curie Intra European Fellowship (G.A.\ No.\ 654916) within Horizon 2020. We would also like to thank F.\ Weik for useful discussions.

\bibliography{nanoparticle_translocation.bib}

\end{document}